\documentclass{article}

\usepackage{arxiv}

\usepackage[utf8]{inputenc} 
\usepackage[T1]{fontenc}    
\usepackage{hyperref}       
\usepackage{url}            
\usepackage{booktabs}       
\usepackage{amsfonts}       
\usepackage{nicefrac}       
\usepackage{microtype}      
\usepackage{graphicx}
\usepackage[numbers,square]{natbib}
\usepackage{doi}
\usepackage{amsmath}  
\usepackage{amssymb}  
\usepackage{hyperref}
\usepackage{svg}
\usepackage{multirow}
\usepackage{caption} 
\usepackage{makecell}

\title{Shifting Attention to You: 

Personalized Brain-Inspired AI Models}

\author{
Stephen Chong Zhao \\
Data Science Institute \\
Vanderbilt University \\
Nashville, United States \\
\texttt{chong.zhao.1@vanderbilt.edu}  \\
   \And
 Yang Hu \\
 Data Science Institute\\
 Vanderbilt University \\
 Nashville, United States \\
  \texttt{yang.hu.1@vanderbilt.Edu} \\
   \And
 Jason Lee \\
 Department of Computer Science\\
 Vanderbilt University \\
 Nashville, United States \\
  \texttt{jason.j.lee@vanderbilt.edu} \\
     \And
 Andrew Bender \\
  Neurosciences Graduate Program\\
  University of California, San Diego\\
  San Diego, United States\\
  \texttt{abender@health.ucsd.edu} \\
     \And
 Trisha Mazumdar \\
  Department of Computer Science\\
  Vanderbilt University\\
  Nashville, United States \\
  \texttt{trisha.mazumdar@vanderbilt.edu} \\
     \And
 Mark Wallace \\
  Department of Psychology\\
  Vanderbilt University\\
 Nashville, United States \\
  \texttt{mark.wallace@vanderbilt.edu} \\
     \And
 David A. Tovar \\
  Department of Psychology\\
  Vanderbilt University\\
  Nashville, United States \\
  \texttt{david.tovar@vanderbilt.edu} \\}

\date{}

\renewcommand{\headeright}{}
\renewcommand{\undertitle}{}
\renewcommand{\shorttitle}{Personalized Brain-Inspired AI Models}

\hypersetup{
pdftitle={Shifting Attention to You: Personalized Brain-Inspired AI Models},
pdfsubject={q-bio.NC, q-bio.QM},
pdfauthor={Stephen Chong Zhao, Yang Hu, Jason Lee, Andrew Bender, Trisha Mazumdar, Mark Wallace, David A. Tovar},
pdfkeywords={Personalization, Artificial Intelligence, Cognition, Transformers, Magnetoencephalography},
}

\begin{document}

\maketitle

\begin{abstract}
The integration of human and artificial intelligence offers a powerful avenue for advancing our understanding of information processing, as each system provides unique computational insights. However, despite the promise of human-AI integration, current AI models are largely trained on massive datasets, optimized for population-level performance, lacking mechanisms to align their computations with individual users’ perceptual semantics and neural dynamics. Here we show that integrating human behavioral insights and millisecond scale neural data within a fine tuned CLIP based model not only captures generalized and individualized aspects of perception but also over doubles behavioral performance compared to the unmodified CLIP baseline. By embedding human inductive biases and mirroring dynamic neural processes during training, personalized neural fine tuning improves predictions of human similarity judgments and tracks the temporal evolution of individual neural responses. Our work establishes a novel, interpretable framework for designing adaptive AI systems, with broad implications for neuroscience, personalized medicine, and human–computer interaction.
\end{abstract}


\section{Introduction}

Modern Artificial Intelligence (AI) systems, much like the human brain, emerge from a complex interplay of model architecture, learning rules, objective functions, and training datasets \citep{goodfellow2016deep, saxe2021deep}. Yet, despite advances in these components, there remains a persistent gap between AI’s computational processes and the nuanced flexibility inherent in human cognition \citep{ferrante2024neuralfoundationmodelsvision, lu2024achieving, rane2024conceptalignment}. Humans effortlessly operate across multiple levels of semantic abstraction, fluidly transitioning between broad categorizations and nuanced distinctions, a capability that even advanced AI models struggle to replicate \citep{goldstone_role_1994,SchulteImWalde2022,born2024evaluating}. Closing this gap by aligning AI more closely with human cognition not only enhances model performance and interpretability but also transforms these systems into powerful tools for probing diverse perceptual processes. To fully harness these benefits, however, AI models must be aligned at the individual level. By capturing the unique cognitive signatures of individual users, personalized AI models offer unprecedented insights into the subtle differences in human perception and provide a novel avenue for understanding the inner workings of individual minds.

While shared architectural principles between deep neural networks (DNNs) and human cortical processing \citep{cichy_comparison_2016} facilitate some degree of brain alignment, recent studies highlight that human-AI alignment may depend more critically on training methodologies than on model architecture itself \citep{muttenthaler_human_2023}. Specifically, the choice of training components, such as the objective function significantly influences the learning of human-like invariances \citep{nanda_invariances_2023}. One promising path is therefore to explicitly incorporate human inductive biases into AI systems, guiding their learning toward cognitive patterns found in human perception. Prior research has leveraged Bayesian model selection and contrastive learning to demonstrate that infusing these biases can enhance both human alignment and general machine learning performance \citep{Sinz2019,marjieh_universal_2023,marjieh_using_2024}.

To directly bridge this gap and build on insights that training methodologies critically shape human-AI alignment \citep{muttenthaler_human_2023,nanda_invariances_2023}, we fine-tuned the Contrastive Language–Image Pre-training (CLIP) model \citep{radford_learning_2021} using human-derived behavioral embeddings generated from large-scale perceptual decision-making datasets \citep{hebart_things_2019,hebart_things-data_2023}. These behavioral embeddings encapsulate interpretable, generalized dimensions of human mental representations, enabling the model to predict human behavioral judgments and neural activity more accurately, significantly surpassing conventional alignment benchmarks \citep{fu2023dreamsimlearningnewdimensions}.

However, while behavioral embeddings capture generalized perceptual dimensions, neural data provide a richer and more granular window into human cognition, revealing millisecond-scale temporal dynamics \citep{Ruffle_2024}. Neural data also uniquely allow researchers to more deeply investigate individual variability, especially valuable in populations where behavioral measurements are impractical or challenging, such as infants or non-verbal individuals \citep{kucyi2024individual}. To leverage these benefits, we developed a dynamic neural fine-tuning method (CLIP-HBA-MEG), aligning our models with rapidly evolving neural patterns measured via magnetoencephalography (MEG).

By first confirming the robustness of our model at the group-level across diverse stimuli and conditions \citep{hebart_things-data_2023}, we extended our approach to fine-tune AI models at the individual participant level. This personalization step allowed us to precisely capture unique neural signatures and cognitive styles, bridging artificial and biological cognition in unprecedented ways. As these personalized models embed distinctive individual perceptual patterns, they not only improve AI’s interpretability and performance but also offer transformative tools for investigating individual differences in cognition. Ultimately, this personalized framework carries deep implications across medicine, cognitive neuroscience, adaptive human-computer interfaces, and tailored AI development, enriching our understanding of both artificial intelligence and the diverse landscape of the human mind.

\section{Results}

\begin{figure}[ht]
    \vspace{-10mm}
    \centering
    \includegraphics[width=\textwidth]{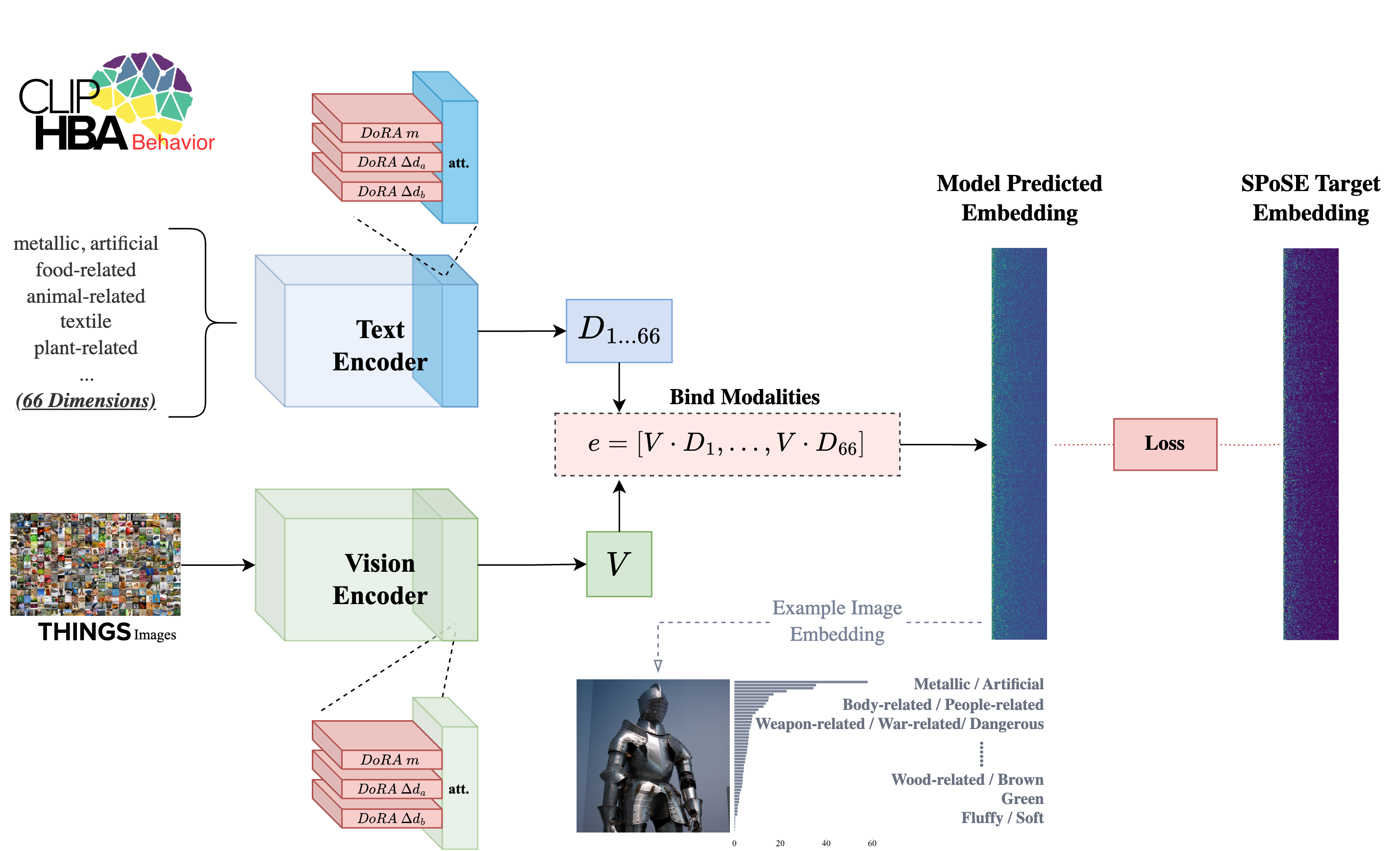}
    \caption{\textbf{Schematic of the CLIP-HBA-Behavior fine-tuning process using behavioral data.} The 66 SPoSE text dimensions are fed into the text encoder, producing 66 text representations $D_1 \dots D_{66}$. Concurrently, visual stimuli from the THINGS dataset are input into the vision encoder, generating their corresponding visual representations $V$. These features from two modalities are bound via a dot product projection, mapping the visual features onto each of the text dimensions to form a 66-dimensional embedding $e$ for each image. Weight-Decomposed Low-Rank Adaptation (DoRA), a parameter efficient fine-tuning (PEFT) method, is used to fine-tune the attention modules of the text and vision encoder, using SPoSE behavioral embedding as an objective.}
    \label{fig:figure1}
\end{figure}

Current AI models trained on large-scale image datasets develop inductive biases based on the statistical patterns within those datasets. However, a significant gap exists between machine-derived inductive biases and the way humans develop theirs through perceptual experience \citep{GoyalBengio2022}. To address this, we introduced human-like inductive biases to the CLIP Model in the form of 66 Sparse Positive Similarity Embedding (SPoSE) dimensions, derived from large-scale behavioral studies \citep{zheng2019revealinginterpretableobjectrepresentations}. These dimensions have been shown to effectively predict human similarity judgments for object distinctions \citep{hebart_revealing_2020}. 

We developed CLIP-HBA-Behavior (Figure \ref{fig:figure1}) by fine-tuning the CLIP model, using the 66 SPoSE dimensions as textual inputs while simultaneously feeding THINGS images \citep{hebart_things_2019} as visual inputs. The CLIP architecture encodes both modalities—textual and visual—separately, before binding them via dot product similarity. This mechanism evaluates the encoded visual features against each of the 66 dimensions, resulting in a 66-dimensional embedding representation for each image input. 

During fine-tuning, the model's predicted embeddings are evaluated against the behavioral SPoSE embeddings using a Mean Squared Error (MSE) loss function. This loss guides the optimization process, updating the attention layers of both the text and visual encoders over multiple training iterations. The goal is to enable the model to accurately predict the SPoSE embedding for a given visual stimulus. In contrast to an off-the-shelf CLIP model, our behaviorally fine-tuned version, CLIP-HBA-Behavior, can produce significantly more human-aligned perceptual outputs. For example, when presented with an image of metal armor (Example Image Embedding in \ref{fig:figure1}), our model identifies the object as highly associated with dimensions such as "metallic," "body-related," and "weapon-related" (top three dimensions), while correctly assigning low relevance to dimensions like "wood-related," "green," or "fluffy" (bottom three dimensions). See \ref{fig:supp1} for more examples.

\subsection{Fine-Tuning CLIP with Human Behavioral Embeddings Enhances Alignment}

\begin{figure}[ht]
    \centering
    \includegraphics[width=\textwidth]{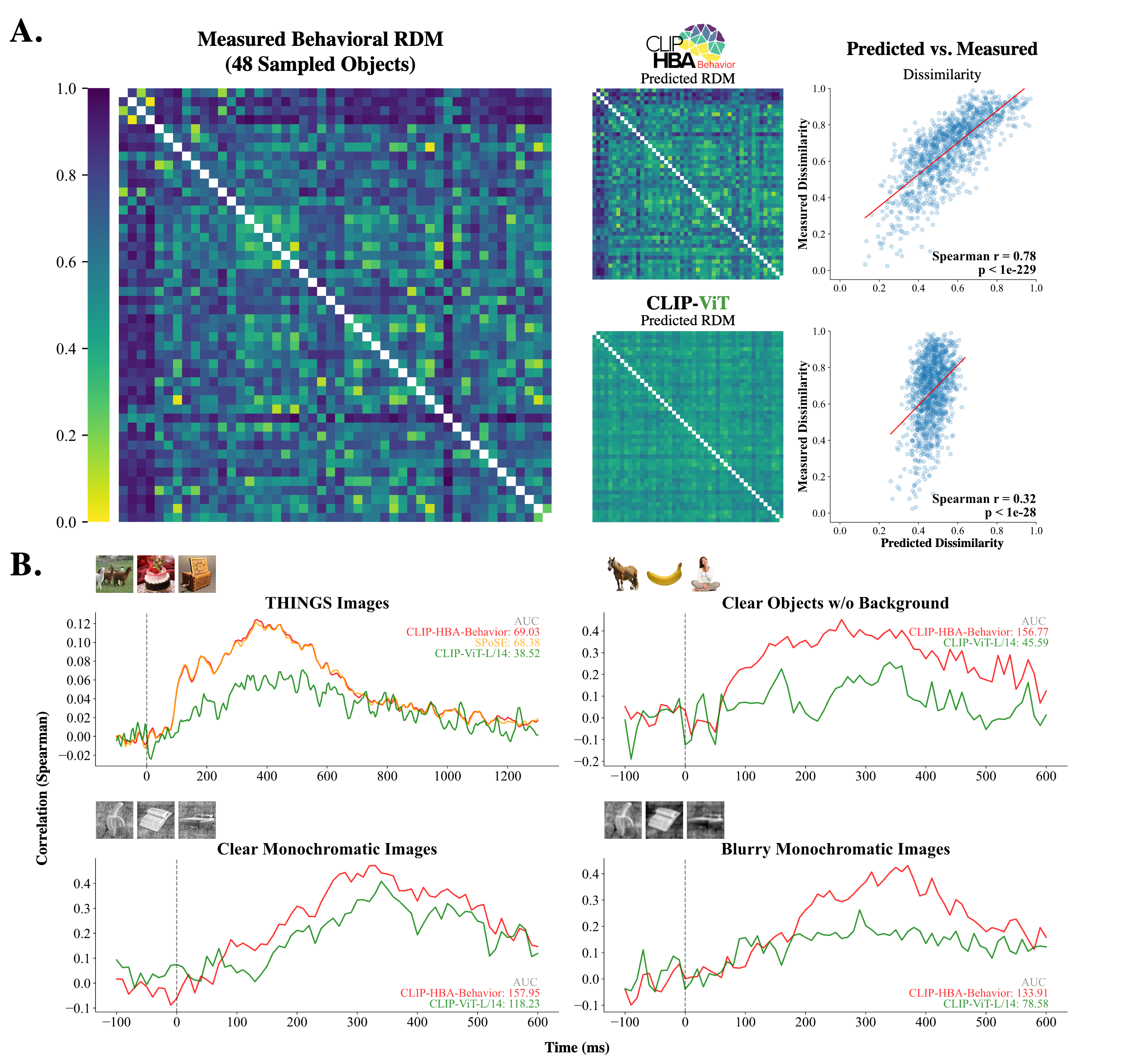}
    \caption{\textbf{Behavioral and neural alignment of fine-tuned CLIP-HBA-Behavior.} (A) Behavioral results: Representational dissimilarity matrices (RDMs) for 48 objects, predicted by CLIP-HBA-Behavior (top) and CLIP-ViT (bottom), with Spearman rank correlations ($\rho$) of 0.78 and 0.32, respectively. (B) Neural results: Temporal correlations between model-predicted RDMs and MEG RDMs from THINGS (top left) and external datasets under varied conditions (remaining panels). CLIP-HBA-Behavior consistently outperforms CLIP-ViT in correlation and area under the curve (AUC), demonstrating enhanced neural alignment and generalizability.}
    \label{fig:figure2}
\end{figure}

As shown in Figure \ref{fig:figure2}A, we evaluated the behaviorally fine-tuned model's performance on a set of 48 object images from the THINGS database, which were fully sampled in a behavioral odd-one-out task \citep{hebart_revealing_2020}. These 48 objects were excluded from training to prevent data leakage. To assess the alignment between the model's predictions and human behavior, we calculated the Spearman rank correlation ($\rho$) between the model's predicted Representational Dissimilarity Matrix (RDM) \citep{kriegeskorte_representational_2008} and the ground-truth behavioral RDM. The behaviorally fine-tuned CLIP-HBA-Behavior achieved a Spearman correlation of 0.78 (\( p<10^{-229} \)) and a 95\% confidence interval (CI) of [0.75, 0.80]. 

For comparison, we applied the same evaluation procedure to the original CLIP-ViT-L14 model. We extracted the 768-dimensional last-layer activations as the model's best representation for each visual stimulus. The Spearman rank correlation between the CLIP-ViT model's predicted RDM and the fully sampled behavioral RDM was 0.32, with a 95\% CI of [0.26, 0.37]. This result highlights the significant improvement achieved by the CLIP-HBA-Behavior model, demonstrating over 100\% improvement in generalizability and behavioral alignment compared to the baseline.

\subsection{Improved Performance on Alignment Benchmark with CLIP-HBA-Behavior}

\begin{table}[ht]
\centering
\renewcommand{\arraystretch}{1.5} 
\captionsetup{skip=10pt} 
\begin{tabular}{llcc}
\hline
\textbf{Feature Type}      & \textbf{Model Name}         & \textbf{Val Score} & \textbf{Test Score} \\ \hline
\multirow{2}{*}{768-d Visual Features} 
                           & \textbf{CLIP-HBA-Behavior}          & 0.88             & 0.88              \\ 
                           & CLIP-ViT-L/14               & 0.81             & 0.81              \\ \hline
\multirow{2}{*}{66-d SPoSE Dimensions} 
                           & \textbf{CLIP-HBA-Behavior}          & 0.85             & 0.84              \\ 
                           & CLIP-ViT-L/14               & 0.80             & 0.79              \\ \hline
\end{tabular}
\caption{\textbf{Performance benchmarking on the NIGHT triplets dataset. }The table compares validation (Val) and test scores for behaviorally fine-tuned CLIP-HBA-Behavior and the baseline CLIP-ViT-L/14 across two feature types: the 66-dimensional SPoSE semantic embedding space and the 768-dimensional visual features. CLIP-HBA-Behavior demonstrates a notable performance enhancement in generalizing to triplet-out decisions after behavioral fine-tuning. }

\label{tab:nights-benchmark}
\end{table}

To test our model's robustness and performance, we benchmarked it against a state-of-the-art human visual similarity judgment metric—NIGHTS (Novel Image Generations with Human-Tested Similarity), a benchmark of 20,019 image triplets with human-tested perceptual similarity scores \citep{fu2023dreamsimlearningnewdimensions}. 

The results in Table \ref{tab:nights-benchmark} illustrate the significant improvements achieved by behaviorally fine-tuning CLIP-HBA-Behavior. To comprehensively evaluate the model, we tested two feature types. First, we assessed the 768-dimensional last layer activations of the Vision Transformer (ViT) encoder's alignment with human perception.  Second, we assessed the 66-dimensional SPoSE embedding space's perceptual alignment. When using last layer visual features, CLIP-HBA-Behavior achieved validation and test scores of 0.88, surpassing the baseline score of 0.81. With the SPoSE dimensions, the model achieved validation and test scores of 0.85 and 0.84, compared to the baseline scores of 0.80 and 0.79. These results highlight that behavioral fine-tuning enhances human perceptual alignment and establishes CLIP-HBA-Behavior as a robust, generalizable model for nuanced perceptual distinctions.

\subsection{Improved Neural Alignment with CLIP-HBA-Behavior}

While the enhanced behavioral correlation of our model was anticipated due to its training on behavioral data, understanding whether the model's embedding space truly becomes more human-like requires evaluating how training exclusively on behavioral data impacts its alignment with neural representations. Therefore, we further evaluated the alignment between the model's representations against neural data obtained from MEG recordings \citep{hebart_things-data_2023}. 

As shown in Figure \ref{fig:figure2}B (top left), CLIP-HBA-Behavior demonstrated a stronger and more sustained temporal correlation with neural data across time points compared to the baseline CLIP model. Additionally, CLIP-HBA-Behavior's correlation pattern and area under the curve (AUC) closely matched—and slightly exceeded—that of the SPoSE embedding model, whereas the CLIP-ViT baseline exhibited notably lower correlations and AUC.

To evaluate generalizability, we tested CLIP-HBA-Behavior on three external datasets containing different participants' neural responses to out-of-distribution object stimuli under varying visual conditions: 

\begin{itemize}
    \item \textbf{Colored Images without Background} (Figure \ref{fig:figure2}B, top right): Clear object stimuli with backgrounds featuring humans, animals, fruits, and man-made objects \citep{grootswagers2017hyperalignment}. 
    \item \textbf{Monochromatic Clear Images} (Figure \ref{fig:figure2}B, bottom left): Clear, monochromatic images of animate and inanimate objects \citep{10.1162/jocn_a_01068}.
    \item \textbf{Monochromatic Images with Blur} (Figure \ref{fig:figure2}B, bottom right): Blurry, monochromatic images of animate and inanimate objects \citep{10.1162/jocn_a_01068}.
\end{itemize}

In all cases, CLIP-HBA-Behavior outperformed baseline CLIP-ViT in terms of AUC. Notably, CLIP-HBA-Behavior consistently peaked in neural alignment around 300–400 ms after stimulus onset. This temporal pattern likely reflects the model's fine-tuning on behavioral decisions and object semantics, which typically emerge later in processing, in contrast to earlier stages related to primary visual processing. 

Overall, CLIP-HBA-Behavior demonstrated enhanced neural alignment across all datasets, spanning diverse image quality conditions and participant groups. Additionally, the performance benchmarking on the NIGHTS dataset highlights the significant improvements achieved by behavioral fine-tuning, with CLIP-HBA-Behavior outperforming the baseline model across both high-dimensional visual features and semantic embeddings. These results underscore that fine-tuning with human behavioral data not only improves the model's ability to capture human cognitive representations but also establishes its robustness and performance in large-scale perceptual tasks. Additionally, we have found that behavioral fine-tuning using this pipeline can be effective with a behavioral dataset as small as 100 visual stimuli, making this method agile and suitable for specialized datasets of nuanced population groups that are likely more difficult to scale \ref{fig:supp3}.

\subsection{Dynamic Neural Fine-Tuning Captures Millisecond-Level Neural Dynamics}

\begin{figure}[ht]
    \centering
    \vspace{-5mm} 
    \includegraphics[width=\textwidth]{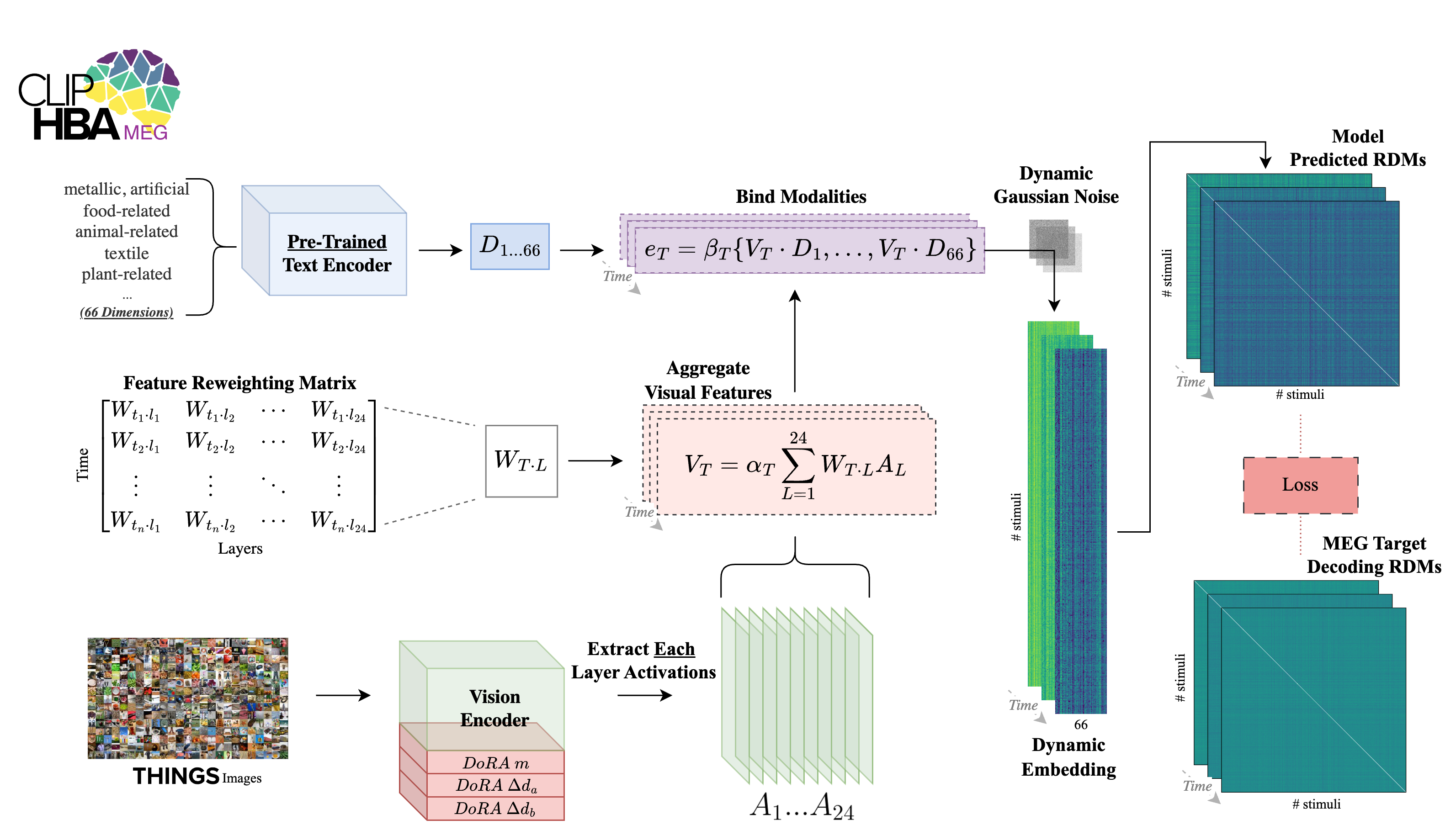}
    \caption{\textbf{Schematic of the CLIP-HBA-MEG fine-tuning process using neural signals.} A Feature Reweighting Matrix, pre-optimized at initialization, dynamically computes weighted combinations of vision encoder layer activations to align with neural decoding RDMs. Temporal scalers, $\alpha_T$ and $\beta_T$, respectively, modulate the magnitude of visual feature aggregation and the binding of visual-semantic features. Dimension-wise Gaussian noise is added to the embedding post-binding with a dynamically controlled noise level, mimicking varying stability of human neural responses while preventing model training from overfitting to specific noisy time points. Predicted RDMs are compared against MEG target RDMs via a custom loss function, updating the vision encoder’s attention layers and adapting the feature reweighting matrix. The text encoder is frozen with behaviorally pre-trained weights to ensure stable semantic features of text representations.}
    \label{fig:figure3}
\end{figure}

Building on the success of perceptual fine-tuning with behavioral data, which are large-scale, generalizable, and collected across many trials and participants, we sought to explore whether the model can improve by learning directly from human neural representations. Although typically collected from smaller participant samples, neural data can often be better suited to capture diverse neural profiles across different populations \citep{Ye_2021}. While behavioral embeddings provide static and generalizable perceptual representations, neural data offer dynamic insights, reflecting millisecond-level temporal patterns of human brain activity that cannot be captured through behavioral measures alone. Furthermore, neural data have the potential to better account for individual variability, providing a more personalized lens into how visual stimuli are processed \citep{Pagan2024}. To explore these possibilities, we developed a dynamic neural fine-tuning process using MEG signals, as demonstrated in Figure \ref{fig:figure3}. The dynamically fine-tuned model, referred to as CLIP-HBA-MEG, adapts its visual stimulus embeddings to align with distinct stages of visual processing and the emergence of semantic representations.

\subsection{Enhanced Behavioral and Neural Alignment after Dynamic Neural Fine-tuning}

\begin{figure}[ht]
    \centering
    \includegraphics[width=\textwidth]{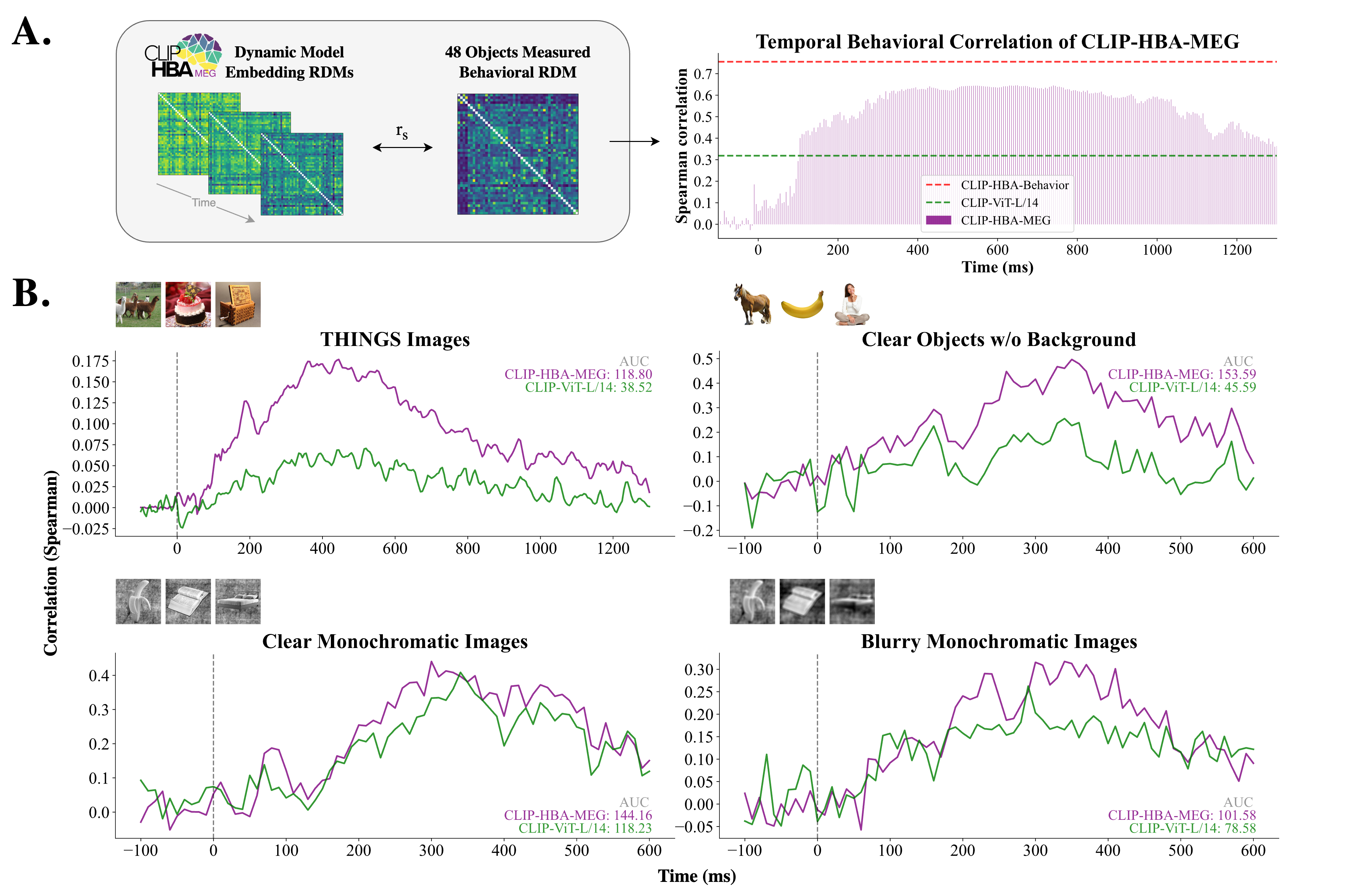}
    \caption{\textbf{Behavioral and neural alignment of fine-tuned CLIP-HBA-MEG.} (A) Behavioral validation: Comparison of the dynamic embedding space of the CLIP-HBA-MEG model across all timepoints (purple bars) with the THINGS behavioral data of 48 sample objects. The static behavioral alignment of the baseline CLIP-ViT model (green line) and the behaviorally fine-tuned CLIP-HBA-Behavior (red line) from Figure \ref{fig:figure2} are also plotted. CLIP-HBA-MEG demonstrates sustained higher behavioral alignment than the baseline soon after stimulus onset but peaks below the alignment achieved by CLIP-HBA-Behavior. (B) Neural validations: Temporal correlations between CLIP-HBA-MEG model-predicted RDMs and MEG RDMs are shown for THINGS images (top left panel) and external datasets under varied conditions: clear objects without background, clear monochromatic images, and blurry monochromatic images (remaining panels). CLIP-HBA-MEG consistently outperforms CLIP-ViT in correlation strength and area under the curve (AUC) across all tested conditions.}
    \label{fig:figure4}
\end{figure}

We evaluated the behavioral and neural validity of the CLIP-HBA-MEG model after fine-tuning with MEG responses from participants viewing visual stimuli in the THINGS image database. Figure \ref{fig:figure4}A illustrates the temporal alignment of the dynamic embedding space of our model against the SPoSE behavioral embedding. Behavioral alignment increased post-stimulus onset, peaking around 600 milliseconds. The peak alignment ($\rho = 0.65$) significantly surpasses the baseline CLIP-ViT's static alignment ($\rho = 0.32$, Figure \ref{fig:figure2}A) but remains below the alignment achieved by the behaviorally fine-tuned CLIP-HBA model ($\rho = 0.78$). These results suggest that, while training with neural data does not match the behavioral performance of the model compared to learning directly from behavioral data, it still significantly improves behavioral alignment over the baseline.

To assess the model's robustness and generalizability in neural alignment, we tested it on the same three external visual datasets with varying characteristics and subjects used in Figure \ref{fig:figure2}B.

As shown in Figure \ref{fig:figure4}B, the neurally fine-tuned CLIP-HBA-MEG model with its dynamic embedding achieved higher overall neural alignment across all tested datasets, presenting higher peak alignment and overall AUC, compared to baseline CLIP-ViT's static representation.

Overall, these findings confirm that the CLIP-HBA-MEG model not only maintains effectiveness in behavioral alignment, but also generalizes effectively across diverse visual conditions, aligning well with human perception and neural responses. However, we observed that the model's neural alignment enhancement was less pronounced on degraded image datasets. This limitation likely stems from the neural fine-tuning process, which exclusively involved clear, naturalistic images from the THINGS database.

\subsection{Viewable Dynamic Attention of CLIP-HBA-MEG}

\begin{figure}[ht]
    \centering
    \includegraphics[width=\textwidth]{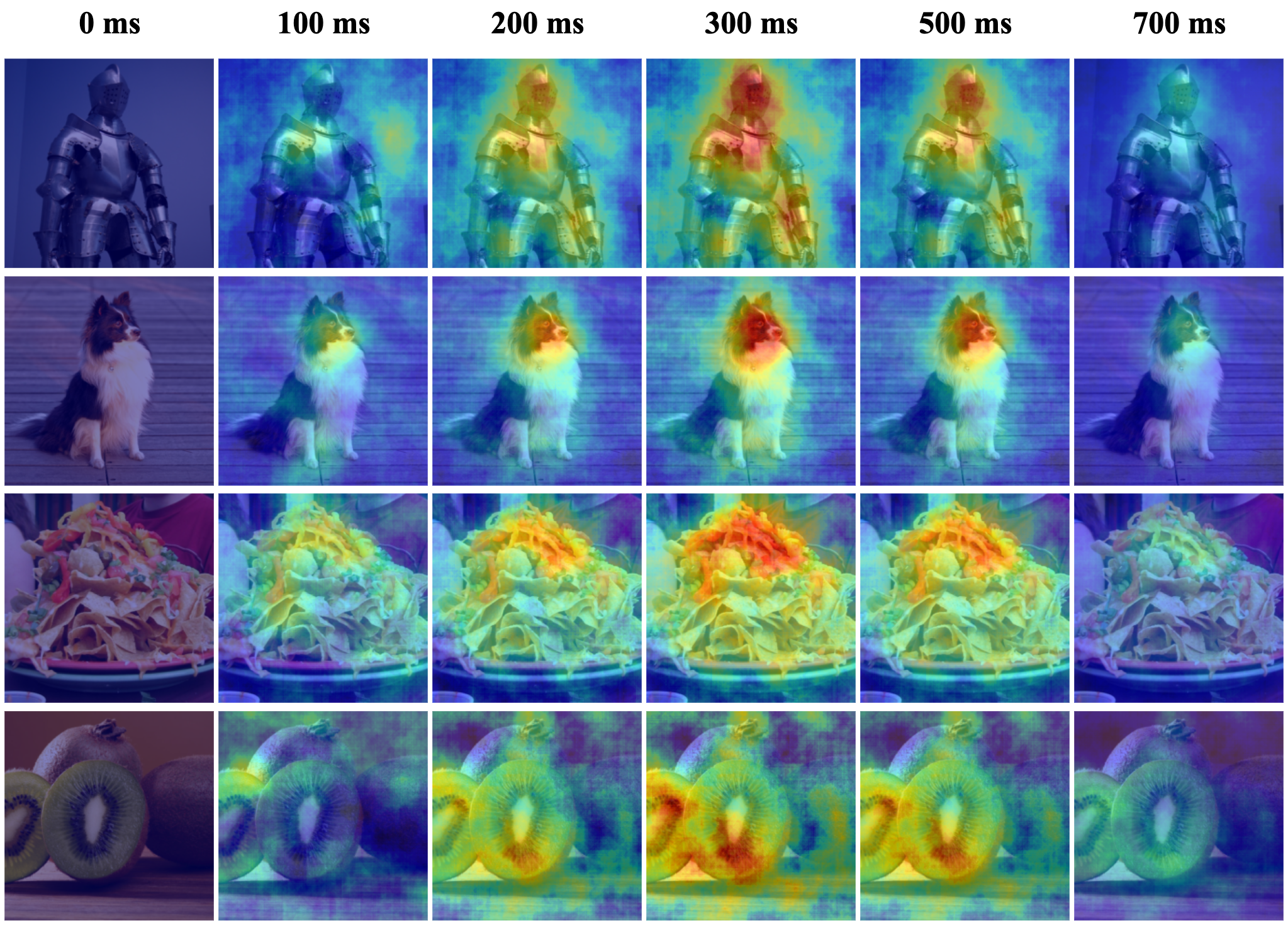}
    \caption{\textbf{Example of dynamic saliency map of neurally fine-tuned CLIP-HBA-MEG}}
    \label{fig:figure5}
\end{figure}

Leveraging the dynamic embedding spaces of the neurally aligned CLIP-HBA-MEG, we can visualize the specific regions and pixels the model prioritizes to produce human-like neural responses at a millisecond resolution. By combining randomized input sampling of the visual stimuli \citep{rise} \citep{kaniuth_high-throughput_2024} with the model’s dynamically learned visual scaler, we can quantify both where and to what extent the model attends within any image following stimulus onset. As shown in Figure \ref{fig:figure5}, the highlighted regions indicate where the model predicts human observers would direct their attention at a given time point. 

\subsection{Individualized Models Capture Participant-Specific Neural Dynamics}

\begin{figure}[ht]
    \centering
    \includegraphics[width=\textwidth]{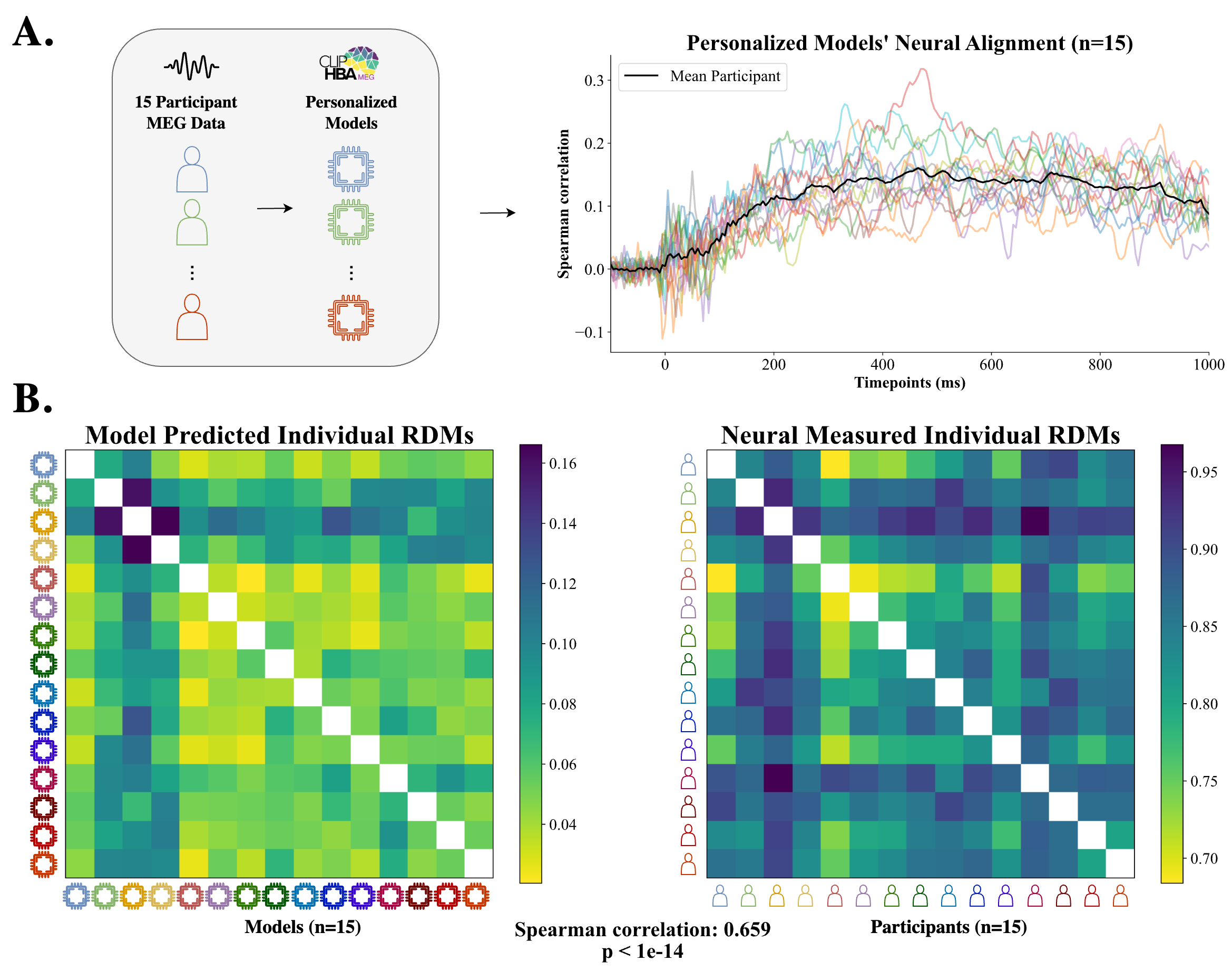}
    \caption{\textbf{Individualized model schematic and results.} (A) Fifteen individualized models were trained on MEG RDMs from 15 participants viewing 100 visual object stimuli. Each model achieved enhanced neural alignment specific to its corresponding participant, demonstrating sustained higher temporal alignment compared to the lower bound starting around 200ms post-stimulus onset. (B) The global correlation (RDMs) between the individualized models and their corresponding participant neural representations was assessed by comparing the differences among fine-tuned personalized models' dynamic representations to the differences among the ground-truth temporal MEG signals of participants. Distances within the personalized models' embedding spaces and participants' MEG RDMs were computed by flattening along the time dimension and measuring Pearson distance between the resulting long vectors. This approach captures both time-specific representations and evolving patterns over time. When assesed using a set of 18 images excluded from training, the Spearman correlation between the RDMs of personalized models and participant neural RDMs was significantly high (\( \rho = 0.659, p < 1 \times 10^{-14} \)), demonstrating that personalized neural fine-tuning successfully captured unique neural dynamics at an individual level.}
    \label{fig:figure6}
\end{figure}

We have demonstrated that our dynamic neural fine-tuning process, trained on group-level MEG data, generalizes well across multiple external datasets and diverse visual conditions, encompassing a wide range of participants and stimuli. These findings underscore the robustness and broad applicability of our approach, reinforcing confidence in its potential. Building on this, we sought to push the boundaries of perceptual fine-tuning by tailoring the model to individual neural profiles. By training on participant-specific MEG signals, we aim to capture unique differences in neural dynamics and signal processing at an individualized level, advancing toward our ultimate goal of creating personalized AI models that reflect the distinct cognitive representations and neural patterns of specific individuals.

As shown in Figure \ref{fig:figure6}, we extended the dynamic neural fine-tuning pipeline of CLIP-HBA-Dynamic to train 15 personalized models, each fine-tuned on the individual MEG data of a single participant \citep{cichy_comparison_2016}. These personalized models achieved enhanced neural alignment specific to their respective participants, exhibiting sustained higher temporal alignment compared to the lower-bound noise ceiling, starting around 200 ms after stimulus onset.

To evaluate whether the personalized models captured consistent individual differences, we assessed the global correlation between the individualized models and the participants’ ground-truth neural representations. This was done by comparing the differences in dynamic embeddings of the fine-tuned models to the differences in temporal MEG signals among participants. Distances within the personalized models' embedding spaces and participants' MEG RDMs were calculated by flattening along the time dimension and computing the Pearson distance between the resulting vectors \ref{fig:supp4}. This approach captured both time-specific neural representations and their evolving patterns over time. Validated on a set of 18 stimuli (excluded from training) viewed by the same 15 participants, the analysis revealed a significant Spearman correlation (\( \rho = 0.659, p < 1 \times 10^{-14} \)) between the RDMs of the personalized models and the participant neural RDMs, confirming that individualized neural fine-tuning effectively captured unique neural dynamics and participant-specific patterns.

\begin{figure}[ht]
    \centering
    \includegraphics[width=\textwidth]{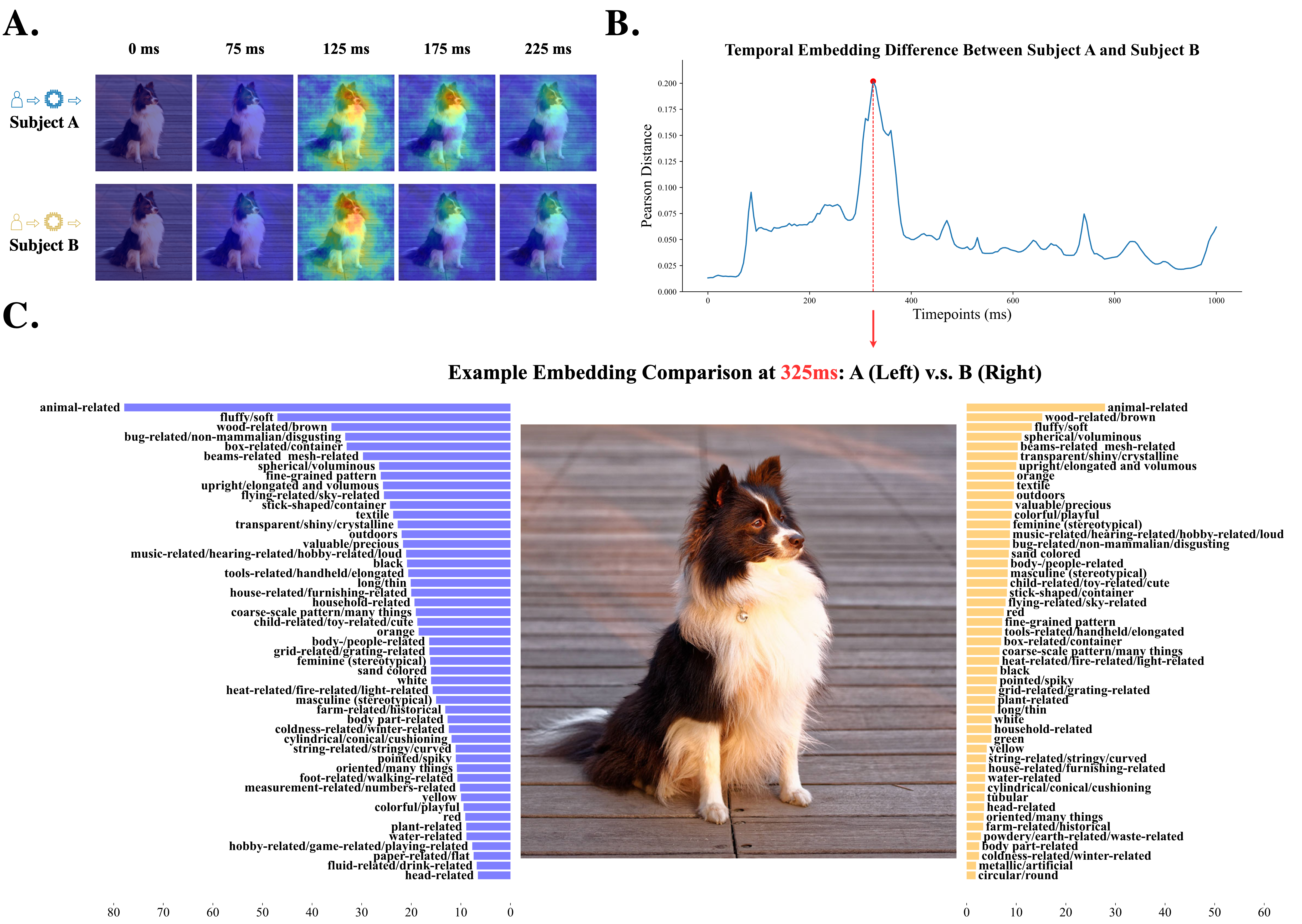}
    \caption{\textbf{Individual Visual and Embedding Differences.} A. Example of individual differences in dynamic attention: Personalized separate CLIP-HBA-MEG models fine-tuned on 2 individual subjects. B. Example of Temporal Embedding Individual Differences: Pearson correlation distance between 2 individual subjects' model-predicted mental embeddings. C. Example of Time-Specific Individual Embedding Differences: 2 corresponding individual models for a given visual stimuli at a time point where their mental embeddings differ the most - 325ms.}
    \label{fig:figure7}
\end{figure}

With the individually aligned models, we can visualize nuanced differences in visual attention and dynamic mental embeddings at the subject level. By applying methods used to analyze model dynamic attention to each personalized model, we observe distinct visual dynamics across subjects following stimulus onset. As shown in Figure \ref{fig:figure7}A, the personalized models of two different subjects exhibit varying temporal attention patterns while both successfully recognizing the presented objects.

Moreover, these personalized models enable the study of how subjects' mental embeddings diverge and converge over time. In Figure \ref{fig:figure7}, the two sample subjects' mental embeddings begin to diverge around 100 milliseconds, peak at 325 milliseconds, and converge after 400 milliseconds. Figure \ref{fig:figure7}C further illustrates dimensional differences in their mental embeddings. At the peak of divergence, Subject A's representation is more skewed, relying on fewer dimensions, whereas dimensions such as "wood-related" rank higher in Subject B's mental embedding. This suggests that at this specific timepoint for the example stimuli, Subject B is more likely integrating background context, while Subject A focuses primarily on the central object. Overall, these findings demonstrate that personalized models offer a powerful tool for studying cognitive processes at a millisecond-level resolution, providing insights into nuanced subject-specific variations.

\section{Discussion}

In this study, we have demonstrated that fine-tuning CLIP with human-derived behavioral and neural embeddings significantly enhances its alignment with human perception and opens the door for individualized AI models. CLIP-HBA-Behavior, trained on large-scale behavioral embeddings, showed a substantial improvement in predicting human similarity judgments while also indirectly capturing neural representational structure. When further refined with millisecond-level MEG data, CLIP-HBA-MEG achieved even greater alignment with human neural dynamics, successfully modeling the temporal evolution of perceptual representations. Notably, the model generalized well across external datasets with diverse participants, visual conditions, and object categories, reinforcing its robustness beyond the training set. Having established its capacity for human-aligned perception at the group level, we then adapted CLIP-HBA-MEG to capture individual differences in neural processing. By fine-tuning models on participant-specific neural data, we found that CLIP-HBA-MEG not only maintained strong alignment with group-level representations but also learned individualized neural dynamics, offering a promising path toward truly personalized AI systems.

\subsection{Scalable and Transferable Applications}

Building on these encouraging findings, our first area of focus was the scalability and transferability of the CLIP-HBA framework. CLIP-HBA-MEG leverages dynamic neural signals to capture real-time perceptual processes and higher-order conceptual organization, offering a nuanced window into human perception that static models cannot provide. In contrast to models that focus primarily on behavioral proxies—such as those that linearly map behavioral embeddings into model output \citep{kaniuth2024high} or perform generalizable local and global transforms to predict human similarity ratings \citep{muttenthaler2023improving}—CLIP-HBA-MEG establishes a direct relationship with neural activity by explicitly capturing the emergence of information from dynamic neural signals. Recent research has shown that even when behavioral outcomes converge, the underlying neural dynamics can differ substantially between individuals \citep{RENART2014211, Bansal_2021}. Such findings underscore the critical importance of capturing dynamic neural activity. While earlier models—such as ReAlnet \citep{lu2024achieving}—attempt to capture cross-modal EEG-fMRI alignments, they lack the temporal resolution necessary to track moment-to-moment perceptual shifts. Moreover, the CLIP-HBA framework can be fine-tuned sequentially, influencing the entire stream of information processing, which uniquely positions it as both a robust model of human perception and a versatile platform that can integrate diverse types of human behavior (e.g., MEG, MRI, behavioral data, electrophysiology). This adaptability is particularly useful in neuroscience applications, where neural data is often limited in scale.

\subsection{Explainable Human-Driven AI Systems}
In addition to capturing real-time perceptual processes, CLIP‐HBA provides fully interpretable dimensions within the model embedding space \citep{hebart_revealing_2020} that are absent in other models designed to incorporate human inductive biases \citep{lu2024achieving}. By integrating this embedding space with dynamic saliency mapping, we offer millisecond‐level insights into both the model’s visual focus and the organization of its internal representations. Furthermore, findings indicate that when model decisions are transparent and systematically quantifiable, it becomes easier to evaluate the consistency of predictions, identify potential biases, and isolate errors for correction \citep{doshi2017towards, rudin2019stop}. These results support the view that interpretability is essential in applications where accurate, accountable decision-making is critical. Together, these advancements move AI closer to human-like cognition by promoting systems whose internal operations are more accessible and amenable to rigorous oversight \citep{NeuroAI2023, Sinz2019}.

\subsection{Expandable Architectures and Modalities for Future Directions}

While our study demonstrates the effectiveness of fine-tuning CLIP to align AI models with human representations using behavioral and MEG data, our approach is not inherently limited to a specific model architecture, type of neural data, or sensory modality. The key factors contributing to the success of our method are the carefully designed loss functions and the modified mechanisms, such as feature reweighting \citep{kaniuth_feature-reweighted_2022} and noise injections. These techniques are broadly applicable to other transformer-based and deep neural network architectures, ensuring that our framework remains adaptable beyond CLIP. This allows our approach to be well-suited for expanding to additional sensory modalities. While our current work focuses on vision, future research could integrate other modalities such as auditory representations for multisensory learning \citep{baltruvsaitis2018multimodal}. By fine-tuning models with neural and behavioral data across multiple sensory domains, we can take the first steps toward a truly multisensory AI system—one that more closely mimics the way the human brain integrates information from diverse sensory inputs \citep{murray2011neural,tovar2020selective}. This expansion would not only enhance AI’s ability to model human perception but also open new possibilities for multisensory cognitive research and human-AI interaction.

\subsection{Applications of Personalized Models}

With the framework's adaptability established, we now turn to its potential for personalization—a critical aspect of aligning AI with individual human cognition. The CLIP-HBA framework represents a significant advancement in personalized AI by integrating human behavioral embeddings with neural data to tailor models to individual cognitive profiles. This approach is deeply rooted in recent cognitive neuroscience research, which suggests that aligning artificial neural networks with biological cognition can enhance both interpretability and robustness \citep{doerig2023neuroconnectionist}. By fine-tuning on millisecond-level MEG signals, the CLIP-HBA-MEG variant directly optimizes neural alignment, reflecting the neuroconnectionist framework’s emphasis on incorporating biological constraints rather than solely focusing on machine-centric performance. Moreover, the framework’s dynamic adaptation of learned representations based on neural feedback mirrors the functionality of digital twin systems \citep{jiang2021industrial}—widely used in industrial and medical applications to model dynamic systems with real-time sensor data—thus acting as a cognitive digital twin that adjusts to an individual’s unique perceptual and neural idiosyncrasies.

Beyond the realm of neuroscience, personalized AI models like CLIP-HBA are making transformative strides in various domains, including education, healthcare, and industry. Advances in AI-driven learning style detection have demonstrated that machine learning can accurately classify users into distinct cognitive subtypes—such as visual, auditory, or kinesthetic learners \citep{kanchon2024enhancing}—which aligns with CLIP-HBA’s ability to fine-tune representations based on individual neural responses. Such personalization is critical for applications where individual differences matter; for example, in clinical contexts, these models can be harnessed to detect subtle deviations in neural dynamics associated with neuropsychiatric conditions \citep{damaraju2014dynamic,preti2017dynamic}, thus informing targeted interventions and enabling clinicians to monitor treatment efficacy over time. Furthermore, the capacity for real-time adaptation paves the way for advanced brain–computer interfaces that adjust to a user’s cognitive state, optimizing interaction and control in assistive technologies \citep{kucyi2024individual}. Collectively, these advancements underscore the transformative potential of personalized AI models to bridge the gap between artificial and biological intelligence, offering adaptive, interpretable, and human-aligned systems across a wide spectrum of fields.

\subsection{Limitations and Need for Diverse Data}

In thinking about the potential applications of personalized models, it is important to recognize the current limitations and the need for more diverse data. The success of our models in aligning with human cognition highlights the need for diverse and representative data. Incorporating behavioral and neural data from a broad participant pool reduces biases from homogeneous training sets, leading to more equitable and generalizable models \citep{born2024evaluating}. However, beyond participant diversity, capturing the variability of real-world sensory experiences is equally important. As demonstrated in Figure \ref{fig:figure2}B, CLIP-HBA-MEG, fine-tuned on group-level MEG data, exhibited strong alignment with human neural representations for clear, naturalistic stimuli but showed reduced generalizability on degraded visual datasets. This suggests that training exclusively on high-quality images may limit practicality and robustness. Since human perception naturally adapts to varied conditions—low lighting, occlusions, and noise—future datasets should reflect this variability by exposing the same participants to a range of visual conditions. This approach would enhance AI’s ecological validity, making models more adaptable and effective in real-world settings.

A major challenge in scaling these efforts lies within the difficulty of collecting large behavioral and neural datasets. While current datasets are often sufficient for traditional studies and comparative analyses, their limitations become increasingly prominent when trying to apply them to AI systems. To address this, exploring data augmentation techniques and leveraging generative AI to create synthetic neural and behavioral datasets could be a promising direction.

\subsection{Conclusion}

Our work underscores the versatility of the CLIP-HBA framework in adapting to a rich spectrum of data modalities, as demonstrated by its robust performance on large-scale, naturalistic stimulus collections. Looking ahead, systematic data collection efforts—ranging from active tasks within scanning environments to naturalistic interactions in immersive settings—hold the promise of revealing how human embeddings evolve when contextual demands change. By exploring active sensing, where representations directly support goal-directed actions, we may uncover entirely new dimensions of cognitive processing. The THINGS initiative \citep{hebart_things_2019, hebart_things-data_2023, hebart_revealing_2020} has already paved the way for training models on diverse data types, and further endeavors that include neurodiverse populations and varied behavioral contexts will only enhance models like CLIP-HBA. Such efforts are not just a testament to the progress made so far but also a compelling challenge to further capture and understand the intricate dynamics of human perception and cognition.

\section{Methods}

\subsection{Participants}

Participants were drawn from existing datasets involving behavioral and neural measurements related to object recognition and mental representations. Behavioral data were obtained from a study that acquired extensive human similarity judgments of natural objects using the THINGS database. \citep{hebart_things_2019,hebart_revealing_2020}. Neural data were sourced from magnetoencephalography (MEG) recordings collected in prior studies \citep{hebart_things-data_2023} \citep{grootswagers2017hyperalignment} \citep{10.1162/jocn_a_01068} \citep{cichy_comparison_2016}. All participants provided informed consent in accordance with institutional guidelines approved by the relevant ethics committees.

\subsection{Materials and Datasets}

\subsubsection{THINGS Dataset and SPoSE Behavioral Embeddings}

The THINGS dataset consists of a comprehensive compilation of 1,854 natural object concepts accompanied by high-quality images and semantic embeddings derived from the Sparse Positive Similarity Embedding (SPoSE) model \citep{hebart_things_2019}. The SPoSE model deduces 66 interpretable human behavioral embeddings based on similarity judgments, capturing the semantic dimensions that structure human mental representations of objects. We subsampled 1806 object stimuli with their corresponding SPoSE embedding for the training and testing of CLIP-HBA-Behavior, excluding 48 fully-sampled stimuli data for measuring the model's behavioral alignment. 

\subsubsection{MEG Data}

We utilized two separate MEG datasets to fine-tune CLIP-HBA-MEG at both group and individual levels (Table~\ref{tab:training_dataset}). For group-level training, we subsampled averaged MEG RDMs from three participants and selected 1,806 out of 1,854 visual stimuli from the THINGS MEG dataset \citep{hebart_things-data_2023}, reserving the same 48 fully-sampled stimuli for behavioral validation. This dataset was chosen for its relatively large stimulus set, ensuring generalizability and providing a robust foundation for demonstrating the model’s ability to learn meaningful neural representations. For individual-level training, we used a dataset with 118 stimuli but a larger sample of 15 participants \citep{cichy_comparison_2016}, prioritizing the number of participants to assess the model's capacity for individualized neural tuning and capturing inter-individual differences. To maintain consistency across training conditions and computational efficiency, we downsampled the individual-level dataset to 200Hz, matching the sampling rate used in group-level training.

\begin{table}[ht!]
\centering
\renewcommand{\arraystretch}{1.5} 
\captionsetup{skip=10pt} 
\begin{tabular}{p{4cm}c cccc}
\hline
\textbf{Usage} & \textbf{Group-Level Training} & \multicolumn{4}{c}{\textbf{Individual-Level Training}} \\ \hline
\textbf{Source} & \citep{hebart_things-data_2023} & \multicolumn{4}{c}{\citep{cichy_comparison_2016}} \\
\textbf{\# Participants} & 3 & \multicolumn{4}{c}{15} \\ 
\textbf{\# Stimuli} & 1854 & \multicolumn{4}{c}{118} \\ 
\textbf{RDM Time Frame} & -100ms -- 1300ms & \multicolumn{4}{c}{-100ms -- 1000ms} \\ 
\textbf{RDM Sampling Rate} & 200Hz & \multicolumn{4}{c}{1000Hz} \\ 
\textbf{RDM Distance Metric} & LDA & \multicolumn{4}{c}{SVM} \\ 
\hline
\end{tabular}%
\caption{\textbf{Summary of Neural Training MEG Datasets}}
\label{tab:training_dataset}
\end{table}

Table~\ref{tab:validation_dataset} includes three external datasets, each with distinct participants, stimuli, and varying visual conditions. These datasets are used to evaluate the neural generalizability of CLIP-HBA-Behavior after behavioral fine-tuning and CLIP-HBA-MEG after group-level neural fine-tuning. By assessing performance on out-of-distribution data, they provide a critical test of the model’s ability to generalize beyond its training set. Both CLIP-HBA-Behavior and CLIP-HBA-MEG are designed to be robust to variations in the sampling rates of validation datasets, adapting dynamically through upsampling or downsampling within the sliding window mechanism to match the target dataset’s sampling rate.

\begin{table}[ht!]
\centering
\renewcommand{\arraystretch}{1.5} 
\captionsetup{skip=10pt} 
\begin{tabular}{p{3.25cm} c c c} 
\hline
\multirow{2}{*}{\textbf{Usage}} & \multicolumn{3}{c}{\textbf{External Validation}} \\ 
\cline{2-4}
& \textbf{Clear/No-Background} & \textbf{Clear/Monochromatic} & \textbf{Blurry/Monochromatic} \\
\hline
\textbf{Source} & \citep{grootswagers2017hyperalignment} & \citep{10.1162/jocn_a_01068} & \citep{10.1162/jocn_a_01068} \\
\textbf{\# Participants} & 20 & 20 & 20 \\ 
\textbf{\# Stimuli} & 36 & 48 & 48 \\ 
\textbf{RDM Time Frame} & -100ms -- 600ms & -100ms -- 600ms & -100ms -- 600ms \\ 
\textbf{RDM Sampling Rate} & 100Hz & 100Hz & 100Hz \\ 
\textbf{RDM Distance Metric} & LDA & LDA & LDA \\ 
\hline
\end{tabular}%
\caption{\textbf{Summary of External Neural Validation MEG Datasets}}
\label{tab:validation_dataset}
\end{table}

\subsection{Model Architecture and Fine-Tuning Procedures}

\subsubsection{Baseline CLIP Model}

We employed the Contrastive Language–Image Pre-training (CLIP) model as our base architecture \citep{radford_learning_2021}. CLIP integrates a visual encoder, built on the Vision Transformer (ViT) architecture \citep{DBLP:journals/corr/abs-2010-11929}, with a text transformer encoder. Pre-trained on a large dataset of image-text pairs, the model aligns visual and textual representations within a shared embedding space. For this study, we specifically employed CLIP-ViT-L/14, a variant with an increased parameter count to enable richer representational learning. This same model configuration serves as our baseline, ensuring a fair comparison when evaluating performance enhancements.

\subsubsection{Fine-Tuning with Behavioral Data}

Figure \ref{fig:figure1} illustrates a schematic of the fine-tuning process for the CLIP model, designed to enhance alignment with human behavior by leveraging SPoSE behavioral embeddings. The model processes inputs from two modalities—image and text—via separate streams. The 66 SPoSE dimensions are tokenized and passed through the text encoder, resulting in 66 textual representations, $D_1 \dots D_{66}$. Simultaneously, image stimuli from the THINGS dataset are processed by the vision encoder, producing corresponding visual representations, $V$. The features from both modalities are then integrated through a dot-product projection, which computes the similarity between the visual features and text dimensions. This process encourages the model to process input stimuli into a human inductive bias mental embedding space. The feature binding function is defined as: 

\begin{equation}
e = \{ V \cdot D_i \}_{i=1}^{66}
\end{equation}

where:
\begin{itemize}
    \item \( \mathbf{e} \) is the model predicted 66-dimensional embedding vector for an input stimulus
    \item \( \mathbf{V}\) is the visual representation of the image stimulus, as produced by the vision encoder.
    \item \( \mathbf{D} = [D_1, \dots, D_{66}] \) is the vector of 66 SPoSE dimensions, as produced by the text encoder.
\end{itemize}

The fine-tuning objective function uses Mean Square Error (MSE) loss, reinforcing the model's predicted embeddings to closely match the SPoSE behavioral target embeddings. The MSE Loss is defined as: 

\begin{equation}
\mathcal{L}_{\text{MSE}} = \frac{1}{N} \sum_{i=1}^{N} \left\| \mathbf{e}^{(i)} - \mathbf{e}_t^{(i)} \right\|^2
\end{equation}

where:
\begin{itemize}
    \item \( e^{(i)} \) is the model predicted 66-dimensional embedding for the input stimulus
    \item \( e_t^{(i)} \) is the target SPoSE 66-dimensional behavioral embedding for the same stimulus
    \item \( N \) is the total number of input stimuli in a training batch.
\end{itemize}

To balance fine-tuning effectiveness with computational efficiency, we employed  Weight-Decomposed Low-Rank Adaptation (DoRA) \citep{liu2024doraweightdecomposedlowrankadaptation} modules into the out-projection of the attention layers, specifically in the final text encoder layer and the last two vision encoder layers. DoRA extends the widely used parameter-efficient fine-tuning (PEFT) method, LoRA \citep{hu_lora_2021}, by decomposing pre-trained weights into separate magnitude and directional matrices. This decomposition enables distinct magnitudinal and directional learning during fine-tuning, closely replicating the optimization dynamics of training directly on the original model weights. Only the DoRA parameters are updated, while the original layers remain frozen, significantly reducing the number of necessary training parameters and computational demands. Furthermore, to leverage CLIP's pre-training on large image-text datasets, we limit the fine-tuning updates to only the last few layers of text and vision encoders, preserving transfer learning capabilities while maintaining the pre-trained model's inherent generalizability \citep{fahes2024finetuningclipsvisualprojector}.

\subsubsection{Behavioral Predictions Evaluation}

We evaluated the fine-tuned CLIP-HBA-Behavior model's ability to predict human behavior using a triplet odd-one-out task on a held-out portion of the data of 48 sample objects. \citep{hebart_revealing_2020}. This set was fully sampled to cover a wide range of naturalistic objects, and these 48 object images were specifically excluded during training to prevent any data leakage. Representational dissimilarity matrices (RDMs) were constructed from the model's embeddings by calculating pairwise cosine distances \citep{kriegeskorte_representational_2008}. These RDMs were compared to RDMs derived from actual behavioral choices using Spearman rank-order correlations \citep{spearman}, assessing the degree of alignment between model predictions and human judgments. 

We evaluated the neural alignment of the fine-tuned CLIP-HBA-Behavior by computing the Spearman rank correlation between the RDM of the model’s static embedding space and the neural average-participant RDM representations at each timepoint. This analysis, conducted using the THINGS dataset and all external validation datasets listed in Table \ref{tab:validation_dataset}, assessed the effectiveness and generalizability of fine-tuning.

\subsubsection{Dynamic Neural Fine-Tuning}

We further developed a dynamic fine-tuning pipeline that enables the model to learn directly from neural representations and dynamics (CLIP-HBA-MEG) using a feature reweighting mechanism. This mechanism dynamically combines intermediate visual features from the vision encoder through a learnable feature reweighting matrix \(\mathbf{W} \) . Figure \ref{fig:figure3} illustrates this pipeline, where \(\mathbf{W} \) operates across temporal and layer dimensions to align model representations with neural dynamics.

The feature reweighing matrix \( \mathbf{W} \in \mathbb{R}^{T \times L} \) where \(T\) is the number of MEG timepoints and \(L=24 \) is the number of visual layers in CLIP's ViT-L/14 encoder. The matrix is initialized such that:

\begin{equation}
W_{t,l} = \begin{cases}
1 & \text{if } l = L \\
0 & \text{otherwise}
\end{cases}
\end{equation}

This initialization strategy assigns full weight to the final layer of the vision encoder while setting all other layers' weights to zero, preserving a CLIP model's pre-trained reliance on the last static representation. During training, \(\mathbf{W} \) is parameterized as a neural network module with constraints enforced through architectural design: each row of \(\mathbf{W} \) is normalized via min-max to ensure non-negative weights summing to 1. 

Optimization follows a two-step gradient descent procedure designed to progressively align the model’s visual processing with human neural representations. Initially, only the feature reweighting matrix is optimized, shifting reliance from static last-layer representations to a dynamic aggregation of intermediate visual features. This enhances temporal dynamics while preserving the pre-trained capabilities of Vision Transformer (ViT) layer weights.

\textbf{Stage 1 - Feature Reweighting:} In this phase, the ViT parameters remain frozen while only the feature reweighting matrix $\mathbf{W}$ is optimized. To enforce temporal consistency, dependencies between neighboring time points are introduced through an average sliding window. Optimization is performed using the AdamW optimizer \citep{adamw} for the initial epochs until convergence. This step establishes foundational temporal layer dependencies without altering the underlying visual processing capabilities of the pre-trained CLIP model.

\textbf{Stage 2 - Joint Fine-Tuning:} Once $\mathbf{W}$ is pre-optimized, all 24 ViT layers join the training process. Joint optimization is performed using DoRA with AdamW optimizer using a new set of learning rate, where $\mathbf{W}$ continues to be updated while the ViT layers undergo refinement. This approach enables the model to learn time-varying feature combinations through $\mathbf{W}$ while refining visual representations towards the target neural data. Detailed training hyperparameters (i.e. learning rates) can be found in Table \ref{tab:hyperparameters}.

The model incorporates temporal modulation through several key components: 1) visual feature magnitude scaling ($\alpha_T$) controlled by neural response richness \citep{Carlson2011} \citep{Carlson2013}, and 2) semantic binding strength ($\beta_T$) determined by time generalization patterns \citep{Tovar2020} \citep{KING2014203}. Additionally, to better mimic the biological nature of human perception—where attention is never fully fixed on a stimulus while ignoring all else—we introduce a mechanism that allows the model to "think of something else" by incorporating dimension-specific Gaussian noise during semantic binding. This ensures a degree of randomness, resembling the natural variability in human neural processing: 

\begin{equation}
\mathbf{e}'_{t} = \beta_T \cdot \left( \mathbf{e}_t + \boldsymbol{\epsilon}_t \odot \sigma_{\mathbf{e}_t} \right)
\end{equation}

where $\mathbf{e}_t$ represents the raw embeddings at time $t$, $\boldsymbol{\epsilon}_t \sim \mathcal{N}(0,1)$ is Gaussian noise, and $\sigma_{\mathbf{e}_t}$ denotes the standard deviation of embeddings across the batch dimension. The noise magnitude is dynamically scaled based on time-dependent neural stability, derived directly from target neural RDMs. This adaptive noise injection allows the model to develop data-driven variability while preventing overfitting to specific noisy time points, such as pre-stimulus recordings or moments long after stimulus onset. By incorporating this mechanism, the pipeline remains robust to variations in the timing of neural recordings, ensuring a more biologically plausible and flexible learning process.

MEG decoding RDMs serve as the target representations within this pipeline. Since neural data lacks a directly comparable 66-dimensional embedding space like the behavioral SPoSE embedding, we transform the model's output embeddings into representational dissimilarity matrices (RDMs) using Pearson correlation \citep{pearson}. This conversion translates the model embeddings into a common representational space with neural data, allowing alignment comparison against the target MEG decoding RDMs.

Given the dynamic nature of the model, its outputs and temporal MEG RDMs are both structured as a 3-dimensional object 

\[
\text{RDM}_{\text{model}}^{\text{object}}, \text{RDM}_{\text{MEG}}^{\text{object}} \in \mathbb{R}^{T \times N \times N}
\]

where: 
\begin{itemize}
    \item $T$ is the number of timepoints
    \item $N$ is the number of batch stimuli
\end{itemize}

This structure reflects the temporal evolution of representations. To construct reliable distance matrices for RDM analysis, a batch size greater than 1 is necessary. Maintaining a batch size of at least $N=32$ is optimal to ensure sufficient RDM sampling, which is critical for the stability of the training process.

A multi-objective loss function was incorporated to align the model's representations with the temporal structure of the neural data. The loss function combines three complementary components: Pearson correlation loss, Mean Squared Error (MSE) loss, and Time Generalization loss.


\textbf{The Pearson correlation loss} ensures structural alignment between predicted and target representations while preserving temporal dynamics. For both model-predicted RDMs and target MEG RDMs, each RDM slice within the three-dimensional temporal RDMs is reduced to its upper triangle and vectorized into a one-dimensional vector (see Supplemental Figure \ref{fig:supp4}). The Pearson loss is then computed as:

\begin{equation}
\mathcal{L}_{\text{Pearson}}\left(x, y\right) = 1 - \frac{\sum_{i} \left( x_i - \bar{x} \right) \left( y_i - \bar{y} \right)}{\sqrt{\sum_{i} \left( x_i - \bar{x} \right)^2} \sqrt{\sum_{i} \left( y_i - \bar{y} \right)^2}},
\end{equation}

where \( x \) represents the vectorized model-predicted RDMs, \( y \) represents the corresponding vector of the target MEG RDMs, and \( \bar{x} \) and \( \bar{y} \) are their respective means.

\textbf{The MSE loss} complements the Pearson loss by regularizing the individual distances between stimuli within compared representations, defined as:

\begin{equation}
\mathcal{L}_{\text{MSE}}\left(x, y\right) = \frac{1}{N} \sum_{i=1}^{N} \left( x_i - y_i \right)^2,
\end{equation}

where \( N \) is the total number of elements in the vectorized RDMs.

\textbf{The Time Generalization loss} regulates the temporal variation and stability of the model's dynamic representation through a multi-step computation. First, we calculate a temporal representational similarity matrix (tRSM) by correlating each RDM slice  with all other remaining timepoints' slices:

\begin{equation}
\text{tRSM}(t_i, t_j) = \frac{1}{N-1} \sum_{k=1}^N \left( \frac{\mathbf{r}_{t_i}^{(k)} - \mu_{t_i}}{\sigma_{t_i}} \right) \left( \frac{\mathbf{r}_{t_j}^{(k)} - \mu_{t_j}}{\sigma_{t_j}} \right)
\end{equation}

where $\mathbf{r}_t$ represents the vectorized upper triangle of the RDM at time $t$, $\mu_t$ and $\sigma_t$ are its mean and standard deviation, and $N$ is the number of stimuli pairs. The generalization profile $g_t$ is then obtained by averaging across the temporal dimension:

\begin{equation}
g_t = \frac{1}{T} \sum_{j=1}^T \text{tRSM}(t,j)
\end{equation}

This captures how neural representations at each timepoint generalize across the temporal evolution. The final Time Generalization loss then measures the alignment between model-predicted ($g^{\text{pred}}$) and neural ($g^{\text{target}}$) profiles through:

\begin{equation}
\mathcal{L}_{\text{Time Generalization}} = \mathcal{L}_{\text{Pearson}}\left(g_t^{\text{pred}}, g_t^{\text{target}}\right)
\end{equation}

where $g_t^{\text{pred}}$ and $g_t^{\text{target}}$ are the respective generalization profiles of the model RDMs and the neural RDMs.

\textbf{The total loss function} for the MEG fine-tuning is:

\begin{equation}
\mathcal{L}_{\text{Total}} = w_1\frac{\mathcal{L}_{\text{Pearson}}}{\mathcal{L}_{\text{Pearson}}^0} + w_2 \frac{\mathcal{L}_{\text{MSE}}}{\mathcal{L}_{\text{MSE}}^0} + w_3 \frac{\mathcal{L}_{\text{Time Generalization}}}{\mathcal{L}_{\text{Time Generalization}}^0},
\end{equation}

where \(\mathcal{L}_{\text{Pearson}}^0\), \(\mathcal{L}_{\text{MSE}}^0\), and \(\mathcal{L}_{\text{Time Generalization}}^0\) represent the initial values of the Pearson loss, MSE loss, and Time Generalization loss, respectively. These values are used as normalization factors to ensure that each loss term is scaled consistently, preventing any single objective from dominating the overall optimization due to differences in magnitude.

\(w_1\), \(w_2\), and \(w_3\) are hyperparameters that weight the contributions of the normalized losses. These parameters need to be adjusted based on different training datasets to ensure balanced optimization, allowing the three objectives to converge smoothly together.

\subsubsection{Neural Data Alignment Evaluation}

We validated the efficacy of our neurally fine-tuned CLIP-HBA-MEG model through both behavioral and neural alignment analyses. For the behavioral alignment evaluation, we used spearman correlation to compare every slice of the model's dynamic RDMs against the static behavioral RDM derived the same 48 sampled stimuli, excluded from the neural fine-tuning process. For the neural alignment evaluation, we assessed the model's performance by comparing its timepoint-to-timepoint RDMs with average-participant MEG decoding RDMs from the THINGS MEG dataset and all mentioned additional validation datasets mentioned in Table  \ref{tab:validation_dataset}, using Spearman rank correlation \citep{spearman}. 

\subsection{Training Individualized Models}

To account for individual differences in neural dynamics, we trained personalized models for each participant using the same CLIP-HBA-MEG pipeline used for group-level training, with a set of adjusted training hyperparameters \ref{tab:hyperparameters} applied to all 15 individualize models. The key distinction was that, instead of using participant-averaged MEG decoding RDMs, each model was fine-tuned specifically on the MEG decoding RDMs corresponding to its respective participant's neural data. 

\subsection{Evaluation of Individual Differences}

Figure \ref{fig:figure6} presents the individual-by-individual dissimilarity matrices for both the neural data and the individualized models. For each pair of participants, we computed the Spearman correlation over time between the model-generated RDMs and the participant’s neural decoding RDMs used during training.

In analyzing the individualized models, temporal RDMs were flattened into one-dimensional vectors for both the model and target. We then computed Spearman correlations between individual models to construct a model-to-model dissimilarity matrix, using the Pearson correlation of the upper triangular elements of the temporal RDMs. Similarly, participant-to-participant RDMs were derived by first computing the Pearson correlation between objects, then flattening the timepoint and upper triangular RDM values into a single vector, as shown in Supplemental Figure \ref{fig:supp4}. Both analyses offer a detailed comparison of representational similarity across participants and their respective models.

Finally, we assessed the correspondence between individualized model RDMs and participant RDMs using Spearman correlation, achieving a strong correlation of 0.65 when evaluated on a left-out validation set of 18 stimuli with the same participants’ neural data.

\subsubsection{Dynamic Saliency Mapping of CLIP-HBA-MEG}
We achieved millisecond-level visualization of dynamically important regions within any given visual stimulus using the fine-tuned CLIP-HBA-MEG model and Randomized Input Sampling for Explanation (RISE) \citep{rise}. The original RISE method identifies critical regions for model processing by comparing classification accuracy between unmasked images and randomly masked inputs.

Since CLIP-HBA-MEG functions as an embedding model without a classification head, we adapted this approach by measuring the cosine similarity between unmasked and masked outputs in the 66-dimensional embedding space. Specifically, if masking a region increases the cosine distance between the masked and unmasked embeddings, that region is inferred to be more crucial for the model’s ability to extract meaningful information and generate effective predicted embeddings. This masking process can be applied to any image at any timepoint within the model’s dynamic embedding space.

Additionally, we incorporated the learned visual scaler ($\alpha$) at each timepoint, allowing for a dynamically evolving visual strength modulated by target neural data. This integration enables us to capture the model’s "dynamic attention" as it processes visual information over time, as demonstrated in Figure \ref{fig:figure5}.

\bibliographystyle{naturemag}
\bibliography{Main}

\clearpage

\section{Supplemental Figures}

\begin{figure}[ht]
    \centering
    \includegraphics[width=\textwidth]{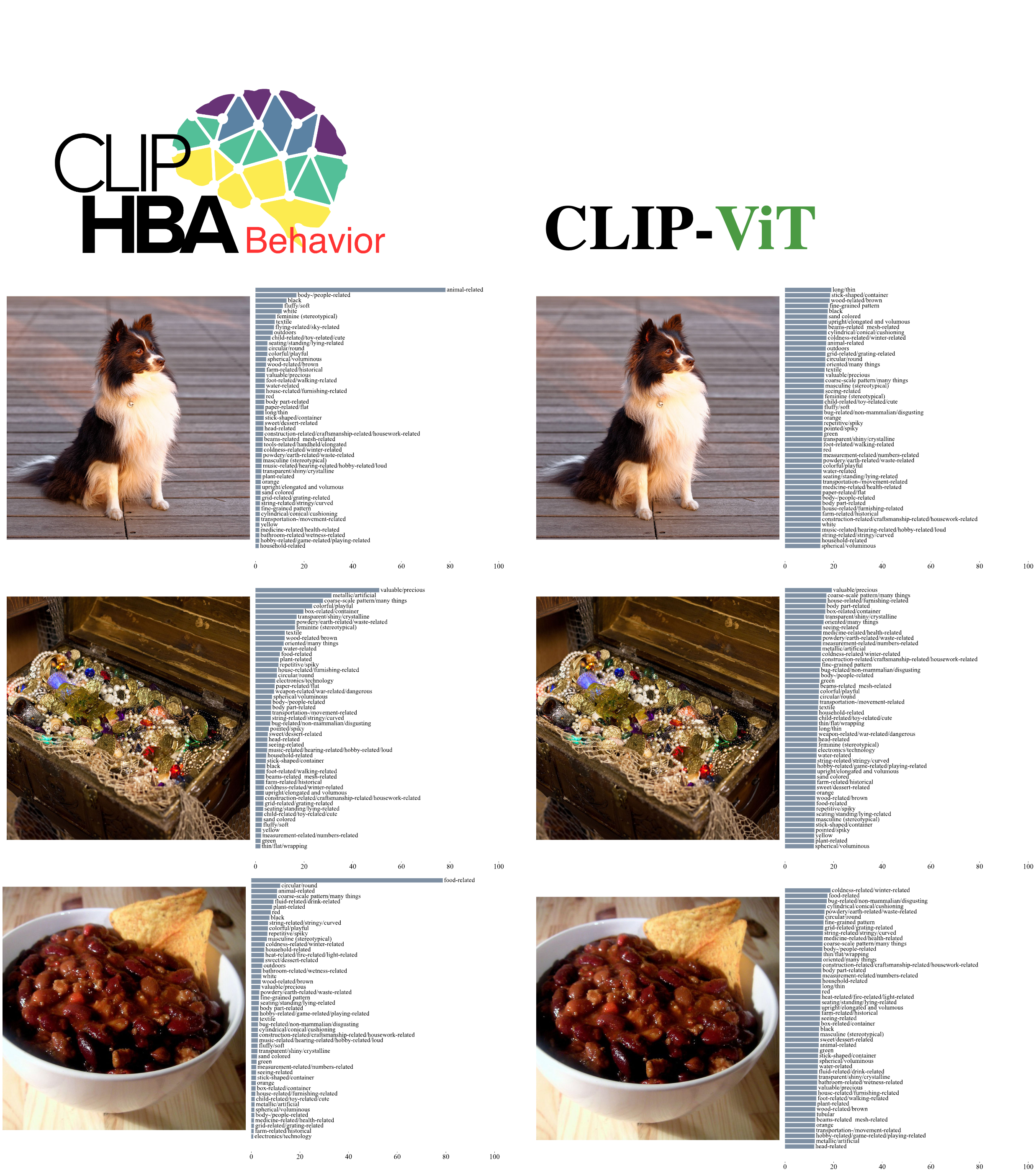}
    \caption{\textbf{Example SPoSE Embeddings of Image Stimuli: }Behaviorally Fine-tuned CLIP-HBA-Behavior v.s. Baseline CLIP-ViT}
    \label{fig:supp1}
\end{figure}

\begin{figure}[ht]
    \centering
    \includegraphics[width=0.55\textwidth]{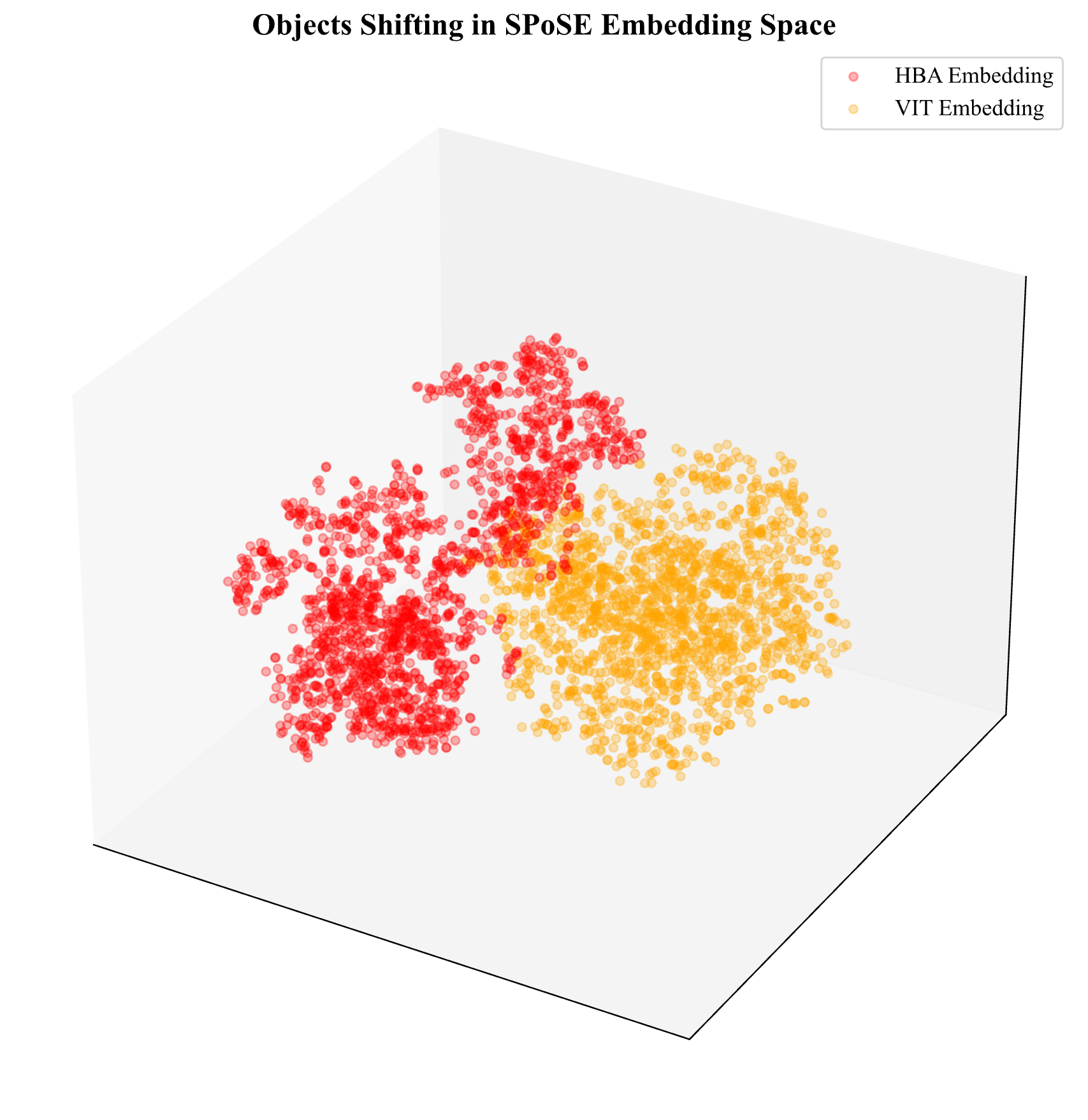}
    \caption{\textbf{THINGS Object Shifting} within the 66-d SPoSE Mental Embedding Space after Behavioral Fine-Tuning; visualized using t-SNE. }
    \label{fig:supp2}
\end{figure}

\begin{table}[ht]
\centering
\renewcommand{\arraystretch}{1.5} 
\captionsetup{skip=10pt} 
\begin{tabular}{>{\centering\arraybackslash}p{6.5cm}llc} 
\hline
\textbf{Models}      & \textbf{Hyperparameters}         & \textbf{Value} \\ \hline
\multirow{8}{*}{CLIP-HBA-Behavior} 
                           & Learning Rate ($\eta$)         & $3 \times 10^{-4}$                 \\ 
                           & DoRA rank ($r$)            & 32                                   \\
                           & DoRA Dropout Rate ($p$)    & 0.1                                   \\
                           & Train Last $n$ Text Layers  & 1                                    \\
                           & Train Last $n$ Vision Layers  & 2                                    \\
                           & Batch Size                    & 64                                    \\
                           & Early Stop Epochs             & 20                                    \\
                           & Train/Test Split              & 8/2                                \\ \hline
\multirow{12}{*}{\makecell{CLIP-HBA-MEG\\(Group-Level)\\(3 Participants, 1806 Stimuli)} }
                           & ViT Learning Rate ($\eta$)         & $3 \times 10^{-5}$                 \\ 
                           & Feature Reweighting $\eta$        & $3 \times 10^{-3}$                  \\ 
                           & Pearson Loss Weight $w_1$     & 1         \\
                           & MSE Loss Weight $w_1$     & 0.1         \\
                           & Time Generalization Loss Weight $w_1$     & 0.15         \\
                           & DoRA rank ($r$)            & 32                                   \\
                           & DoRA Dropout Rate ($p$)    & 0.1                                   \\
                           & Train Last $n$ Text Layers  & 0                                    \\
                           & Train Last $n$ Vision Layers  & 24 (All)                              \\
                           & Batch Size                    & 64                                    \\
                           & Early Stop Epochs             & 5                                    \\
                           & Train/Test Split              & 1456/350                                \\ \hline
\multirow{12}{*}{\makecell{CLIP-HBA-MEG\\(Individual-level)\\(15 Participants, 100 Stimuli)} }
                           & ViT Learning Rate ($\eta$)         & $3 \times 10^{-5}$                 \\ 
                           & Feature Reweighting $\eta$        &  $3 \times 10^{-3}$                 \\ 
                           & Pearson Loss Weight $w_1$     & 1         \\
                           & MSE Loss Weight $w_1$     & 0.15         \\
                           & Time Generalization Loss Weight $w_1$     & 0.1         \\
                           & DoRA rank ($r$)            & 6                                   \\
                           & DoRA Dropout Rate ($p$)    & 0.1                                   \\
                           & Train Last $n$ Text Layers  & 0                                    \\
                           & Train Last $n$ Vision Layers  & 24 (All)                              \\
                           & Batch Size                    & 40                                    \\
                           & Early Stop Epochs             & 10                                    \\
                           & Train/Test Split              & 80/20                                \\ \hline
\end{tabular}
\caption{\textbf{Hyperparameters Used for Model Training.}}
\label{tab:hyperparameters}
\end{table}

\begin{figure}[ht]
    \centering
    \includegraphics[width=0.8\textwidth]{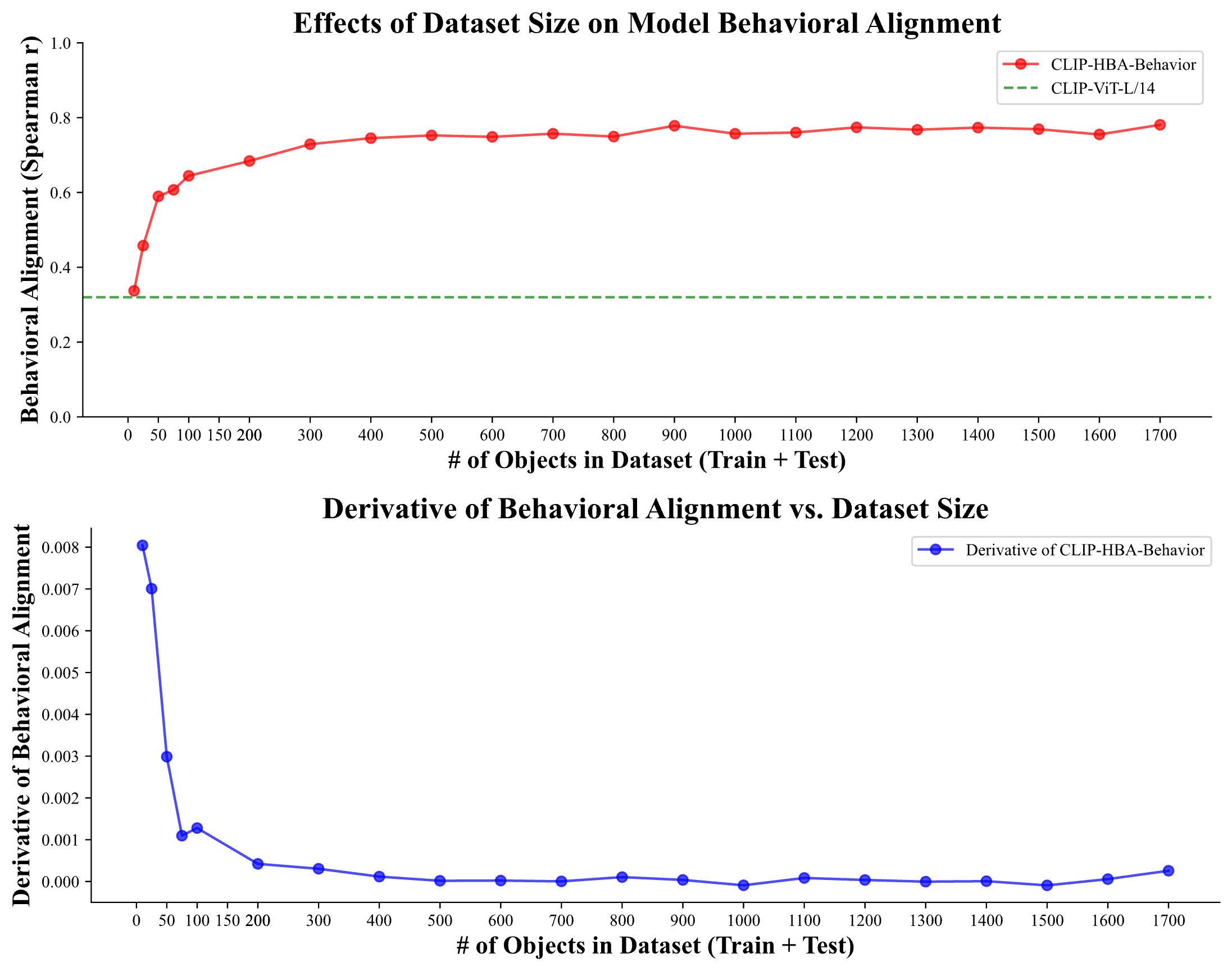}
    \caption{\textbf{Evaluation of Training Effectiveness based on the Size of the Behavioral Dataset}. CLIP-HBA-Behavior models are trained using random subsamples of the 1854-object THINGS behavioral dataset, and evaluated against the 48 excluded validation samples' triple-out behavior ground truth. Enhancement in behavioral alignment starts to diminish after the dataset size surpasses 100 object images. }
    \label{fig:supp3}
\end{figure}

\begin{figure}[ht]
    \centering
    \includegraphics[width=0.8\textwidth]{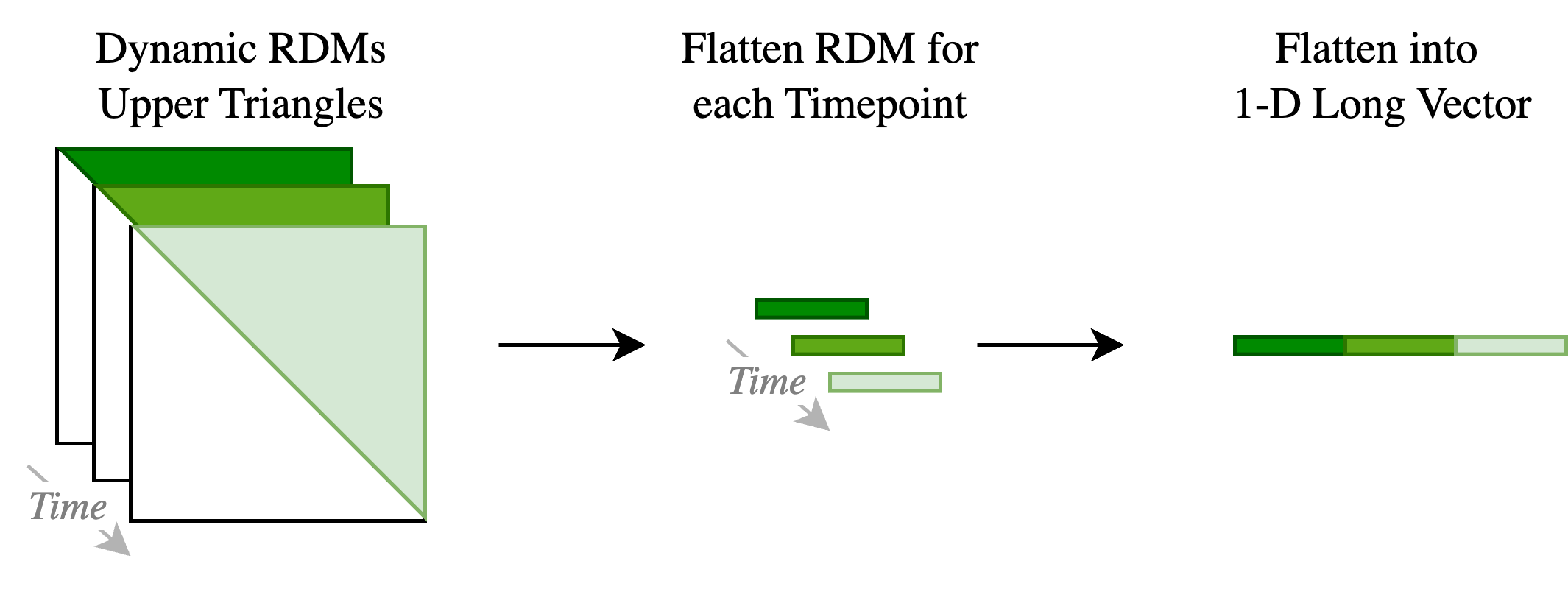}
    \caption{\textbf{Schema of converting dynamic RDMs into a 1-D vector for distance and correlation comparisons.}. Each 3D RDM object represents dynamic representations, with each slice corresponding to a specific millisecond time point. For each time slice, we extract and flatten the upper triangle of the RDM (excluding the diagonal). These flattened slices are then concatenated into a single 1-D vector, preserving temporal dynamics and representational structure for comparison between model representations and human MEG data. }
    \label{fig:supp4}
\end{figure}


\end{document}